\documentclass[iop,onecolumn,preprint,superscriptaddress,showpacs,nofootinbib]{revtex4}
\usepackage{epsfig}
\usepackage{hyperref}
\usepackage{color}
\usepackage{amsmath}
\usepackage{subeqnarray}
\def\laq{\raise 0.4ex\hbox{$<$}\kern -0.8em\lower 0.62ex\hbox{$\sim$}}
\def\gaq{\raise 0.4ex\hbox{$>$}\kern -0.7em\lower 0.62ex\hbox{$\sim$}}
\newcommand{\be}{\begin{equation}}
\newcommand{\ee}{\end{equation}}
\newcommand{\bea}{\begin{eqnarray}}
\newcommand{\eea}{\end{eqnarray}}

\begin{document}
\title{ Extraction of energy from gravitational waves by laser interferometer detectors}
\author{Yiqiu Ma}
\address{1: School of Physics, University of Western Australia}
\author{David G Blair}
\email{david.gerald.blair@gmail.com}
\address{1: School of Physics, University of Western Australia}
\author{Chunnong Zhao}
\address{1: School of Physics, University of Western Australia}
\author{William Kells}
\address{2: California Institute of Technology}
\begin{abstract}
In this paper, we discuss the energy interaction between gravitational waves and the laser interferometer gravitational wave detectors. We show that the widely held view that the laser interferometer gravitational wave detector absorbs no energy from gravitational waves is only valid under the approximation of a frequency-independent optomechanical coupling strength and a pump laser without detuning with respect to the resonance of the interferometer. For a strongly detuned interferometer, the optical-damping dynamics dissipates gravitational wave energy through the interaction between the test masses and the optical field. For a non-detuned interferometer, the frequency-dependence of the optomechanical coupling strength causes a tiny energy dissipation, which is proved to be equivalent to the Doppler friction raised by Braginsky {\it et al.}
\end{abstract}

\maketitle
\section{Introduction}
\label{sec1}
Astronomically large fluxes of gravitational waves are expected to be detected by advanced laser interferometer gravitational wave detectors such as Advanced LIGO and Advanced Virgo now being commissioned \,\cite{Harry2010,Virgo}. For example, a binary black hole coalescence at 1Gpc distance, which has peak luminosity $\sim10^{23}L_\odot$, has a flux at the Earth of $\sim10\text{W}\text{m}^{-2}\text{s}^{-1}$, which vastly exceeds the flux of all electromagnetic astronomical sources except the Sun.
Clearly a large amount of energy is available in the signals. However the extremely weak interaction of gravitational wave detectors with gravitational waves makes detection very difficult and is the main reason that gravitational wave detection has not yet been accomplished.

This paper addresses the question of how much energy can be extracted from gravitational waves. Traditionally only resonant mass gravitational
wave detectors have been understood from an energy interaction viewpoint. Following Weber~\cite{JW}, the sensitivity of resonant mass detectors was estimated by considering the work done by an incident gravitational wave. However for
laser interferometer gravitational wave detectors
estimation has normally been based on considering the test
masses as free masses that experience the gravitational wave spatial
strain $h$ of a passing wave. If the masses
are truly free, no energy is extracted from the wave. The free-mass approximation naturally
leads to an approach that neglects the energy interaction. However, as emphasised by Saulson: ``an important kind of understanding is lost in the neglect of such an essential physical concept as energy"\,\cite{PS}.

The above discussion recalls the debate about the existence of
gravitational waves that occurred from $1916$ to $1957$ ~\cite{DK}.
The proof of the reality of gravitational waves was eventually
clarified by the rubbing sticks gedanken experiment presented by Feynman at the 1957 Chapel Hill Conference\,\cite{DeWitt}. Feynman showed that gravitational waves are able to deposit frictional energy and therefore cannot be a mathematical artifact. This leads to the viewpoint that a practical detector
must be a transducer for gravitational waves, converting wave energy
into electromagnetic energy, and amplifying it to enable it to be
resolved against the inevitable background of instrument noise.

In this paper we discuss the fundamental question of energy
absorption in relation to laser interferometer gravitational wave
detectors by studying the energy flow. Our discussion is designed to illuminate fundamental principles in the context of new and more general interferometer designs, and to present results consistent with the concept of energy absorption cross section. Because laser interferometers operate in the quantum regime it is necessary to use a full quantum optomechanical analysis.

We will begin our analysis with a quantum analysis of the Doppler friction effect which arises from the frequency change of photons on reflection from a moving mirror. This was first discussed by Einstein in a thought experiment ~\cite{Einstein} and then rediscovered by Braginsky etl.al~\cite{Braginsky} and Saulson~\cite{PS}.  Saulson showed that this effect indeed provides a viscous coupling to gravitational waves. While it is like the friction between Feynman's sticks, the effect is small because it is a second order relativistic effect ($\sim (v/c)^2$, in which $v$ is the speed of test mass motion and $c$ is the speed of light. In this paper, we derive this friction from a quantum mechanical viewpoint, and give a classical derivation in the Appendix A. For a typical predicted wave and a LIGO-like interferometer, Saulson showed that the power absorbed is $\sim10^{-40} W$. If the primary gravitational wave signal was provided by this mechanism, the detector would need to have power gain that scales as the square of the ratio of optical frequency to gravitational wave frequency.
However, we show here that the Doppler friction is not the primary signal source but a small and generally negligible additional term that can also be interpreted as a result of unbalanced Stokes and anti-Stokes sidebands. Our analysis of a toy model reveals that the power gain of the detector follows the usual form for parametric transducers, scaling linearly with the frequency ratio.

Through analysis of interferometers we will show that the free mass approximation is indeed an excellent approximation for detectors constructed to date, which all use a balanced pair of sidebands, but that it is not valid for more general detector configurations such as detuned ~\cite{BC} or double optical spring ~\cite{RM} interferometers in which the sidebands are unbalanced. In this case there can be strong optical damping which give rise to much stronger absorption of gravitational wave energy. Finally having demonstrated how energy absorption is related to sideband unbalance, and with view to stimulating new thinking about detectors, we mention the tilt interferometer as an example of a detector which has a single sideband and hence maximal sideband imbalance.~\cite{DGBlair}

The purpose of this paper is to to address the single issue of energy absorption in the context of modern interferometer concepts, and to point out that gravitational wave energy absorption can be engineered into detector designs. We begin by considering a toy model to show that Doppler friction appears naturally as long as the frequency dependence of the optomechanical coupling strength is included. In section 3, we extend the discussion to general interferometer configurations, giving a rigorous derivation of energy dissipation through optical damping, which allows us to derive an energy absorption cross section which is analogous to  that of resonant bar detectors.

\section{ Optomechanical interaction between light and a mirror}
In laser interferometer gravitational wave detectors the basic physical process is the interaction between the light beam and the center
of mass (CoM) degrees of freedom of the mirrors. The CoM motion of the mirrors, driven by
gravitational waves,  modulates the light beam and creates anti-Stokes (upper)
and Stokes (lower) sidebands with frequency $\omega_c\pm\Omega$. Here, $\omega_c$
is the frequency of the carrier beam and $\Omega$ is the frequency of gravitational
waves.

This modulation process can also be treated in the quantum picture as generation of Stokes
and anti-Stokes photons by scattering between the carrier photon and mechanical phonon,
which can be described by the following Feynman diagrams shown in Fig.1

\begin{figure}
\begin{center}
\epsfig{file=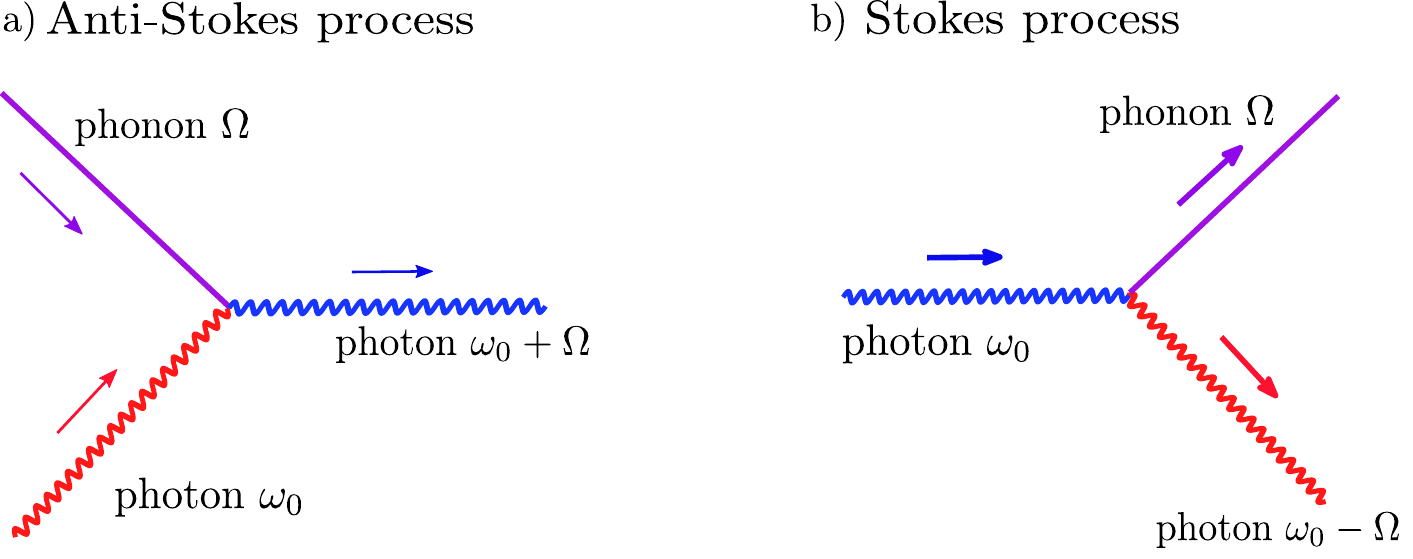,width=0.65\textwidth,height=0.3\textwidth,angle=0}
\caption{Parametric interaction in terms of Feynman diagram, (a) The anti-Stokes process,
which causes a cooling effect by drawing a phonon out of the mechanical degrees
of freedom (mirrors), creating an upper sideband photon with higher energy $\hbar(\omega_0+\Omega)$;
(b) The anti-Stokes process, creates a heating effect by emitting a phonon to
the mechanical degrees of freedom, creating a lower sideband photon with lower energy
$\hbar(\omega_0-\Omega).$}
\label{Fig0}
\end{center}
\end{figure}

From these Feynman diagrams, it is clear that if the rate of the anti-Stokes process is higher than that of the Stokes process, more
phonons will be absorbed through the anti-Stokes process than emitted through the Stokes process.
In this case, there is a net flow of mechanical energy into the light field, and vice versa.

A widely-held view of this modulation process is that the Stokes and anti-Stokes sidebands are
`balanced'.  This means that the creation of an anti-Stokes sideband photon must be accompanied by the creation
of a Stokes sideband photon. In other words, the Stokes and anti-Stokes photon generation rates are equal, implying that there is no net energy transfer between the mirror and the optical field when a free propagating
laser field is modulated by mirror motion.

In this section, by carefully analyzing a toy model, we will show that this viewpoint is only approximately correct. What
has been neglected here is  the Doppler friction discussed by Braginsky and Saulson in
~\cite{Braginsky}~\cite{PS}. We will derive the dynamics of the model, then
give an intuitive interpretation of the result.

\subsection{System Dynamics}
First, we review the derivation of the optomechanical coupling Hamiltonian using a toy model consisting of a light beam reflected by a mirror, see Fig.2. The light beam, which is accompanied by quantum fluctuations, is given by:
\begin{equation}
\begin{split}
E_{in}=2\sqrt{\frac{2\pi\hbar\omega_0}{Sc}}E_0\cos{(\omega_0t)}
+e^{-i\omega_0t}\int^{\infty}_0\frac{d\Omega}{2\pi}(\sqrt{\frac{2\pi\hbar\omega_+}{Sc}}\hat{a}_+e^{-i\Omega t}+\sqrt{\frac{2\pi\hbar\omega_-}{Sc}}\hat{a}_-e^{i\Omega t})\\
+e^{i\omega_0t}\int^{\infty}_0\frac{d\Omega}{2\pi}(\sqrt{\frac{2\pi\hbar\omega_+}{Sc}}\hat{a}^{\dagger}_+e^{i\Omega t}+\sqrt{\frac{2\pi\hbar\omega_-}{Sc}}\hat{a}^{\dagger}_-e^{-i\Omega t}).
\end{split}
\end{equation}
Here, $\omega_0$ is the pumping frequency of the steady part of the incoming light beam,
$E_0$ is the amplitude of steady field, $S$ is its transverse cross-section and $\hat{a}_{\pm}$ ($\hat{a}^\dagger_{\pm}$)
are the annihilation (creation) operators of the optical field
at the sideband frequencies $\omega_{\pm}=\omega_0\pm\Omega$. The first term here is the steady part of the optical field while the second part is
the fluctuating part which has a continuous frequency distribution.

This optical field exerts a radiation pressure force $F=2|E_{in}|^2S/4\pi$
and does work on the mirror.
Therefore the interaction Hamiltonian is: $H=-F\cdot x$.
After substituting (1) and keeping the first order terms, we have:
\begin{equation}
H_{int}=-\frac{\hbar E_0}{c}\int^{\infty}_0\frac{d\omega}{2\pi}\sqrt{\omega_0\omega}(\hat{a}_\omega e^{-i(\omega-\omega_0)t}+
\hat{a}^{\dagger}_\omega e^{i(\omega-\omega_0)t})\cdot x.
\end{equation}
We will neglect the non-interesting steady part $\propto|E_0|^2\hat{x}$  since it can be balanced by exerting an external constant force.
We also rewrite $\hat{a}_{\omega_0\pm\Omega}$ to be $\hat{a}_\omega$. This form of Hamiltonian can also be found in ~\cite{Bra}~\cite{SD}.

It is important to notice that the coupling strength at frequency $\omega$
is now proportional to $ \sqrt{\omega_0\omega}$, a factor that comes from the beating
between the steady and fluctuating optical amplitude. Usually, we treat
$\omega\sim\omega_0$, thereby approximating the optomechanical coupling strength as a
frequency independent constant. However clearly the coupling strength is not frequency independent.

The Heisenberg equations describing the evolution of the mirror-field system are given by:
\begin{subequations}
\begin{align}
&\frac{d\hat{a}_\omega}{dt}=i\frac{E_0}{c}\sqrt{\omega\omega_0}\hat{x}(t)e^{i(\omega-\omega_0)t},\\
&\frac{dp}{dt}=\frac{\hbar E_0}{c}\int^{\infty}_0\frac{d\omega}{2\pi}\sqrt{\omega\omega_0}[\hat{a}^{\dagger}_\omega(t)e^{i(\omega-\omega_0)t}+\hat{a}_\omega(t)e^{-i(\omega-\omega_0)t}],\\
&\frac{dx}{dt}=\frac{p}{m}.
\end{align}
\end{subequations}

First we consider the steady optical field and
 the fluctuating component due to modulation by the mirror motion, neglecting the
 quantum fluctuation field. then the generation of sideband field with frequency $\omega_0+\Omega$ is due to the
 mirror oscillation at frequency $\Omega$. We also assume
 an \emph{initial condition} that at $t=0$, there is no light except the pumping field
 at $\omega_0$. Solving the above Heisenberg
 equations, we have:
\begin{subequations}
\begin{align}
&\hat{a}_\omega(t)=i\frac{E_0}{c}\sqrt{\omega\omega_0}\int^{t}_{t_0}x(t^{\prime})e^{-i(\omega-\omega_0)t^{\prime}}dt^{\prime},\\
&F_{rad}=m\frac{d^2x}{dt^2}=\frac{dp}{dt}=-\frac{2\hbar E_0^2\omega_0}{c^2}\int^{t}_{t_0} dt^{\prime}\int^{\infty}_0\frac{d\omega}{2\pi}[\omega x(t^{\prime})\sin{(\omega-\omega_0)(t^{\prime}-t)}].
\end{align}
\end{subequations}

Equation (4b) can be derived by substituting (4a) into (3b). Substituting $\omega=\omega_0+\Omega$ into (4b), we can separate out
a force term dependent on $\Omega$:
\begin{equation}
F^{\Omega}_{rad}=-\frac{2\hbar E_0^2\omega_0}{c^2}\int^{t}_{t_0}dt^{\prime}\int^{\infty}_{-\infty}\frac{d\Omega}{2\pi}\Omega x(t^{\prime})\sin{\Omega(t^{\prime
}-t)}.
\end{equation}
Integrating by parts, and picking up the velocity dependent term which is related to the energy absorption, we have:
\begin{equation}
\begin{split}
F_{v}=-\frac{2\hbar E_0^2\omega_0}{c^2}\dot{x}(t).
\end{split}
\end{equation}
in which $\dot{x}(t)$ is the velocity of the test mass motion.
Following the logic of the above argument, we can easily see that if we impose the approximation $\omega\sim\omega_0$, we will not have the $\Omega$-dependent radiation pressure term as in Eq.(5), and hence no velocity dependent force.
\begin{figure}
\begin{center}
\epsfig{file=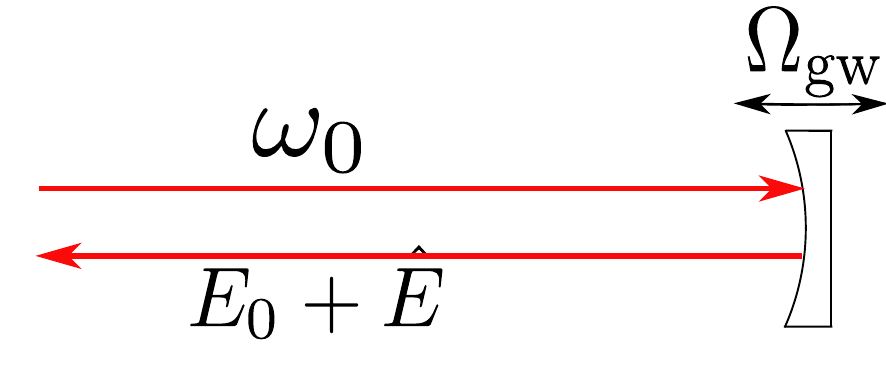,width=0.4\textwidth,height=0.2\textwidth,angle=0}
\caption{Toy model for Doppler friction: The
optical field is modulated by an oscillating mirror with frequency $\Omega_{\rm gw}$. Because of this modulation, the optical field consists of two parts, namely, the pumping part $E_0$ and the sideband part $\hat E$.}
\label{Fig0}
\end{center}
\end{figure}

Suppose the free test mass is driven by a monochromatic gravitational wave with frequency $\Omega_{\text{gw}}$ and strain $h$, then we have $\dot{x}(t)=\Omega_{\rm gw}hL\cos(\Omega_{\rm gw}t)$ in the steady state. Then Eq.(6) can be written as
\be
F_v(t)=-\frac{2P\Omega_{\text{gw}}}{c^2}hL\cos(\Omega_{\rm gw}t)
\ee
in which $L$ is the distance from the equilibrium position of the mirror to a reference point. The power is given by $P=\hbar\omega_0|E_0|^2$ and the motion is at frequency $\Omega_{\text{gw}}$. Therefore Eq. is exactly the Doppler friction force given in ~\cite{PS}. In Appendix A, we give a classical derivation that gives the same result. For the real interferometer with optical resonant cavities, we only need to multiply the above formula by the folding factor $N_{fold}$ as in ~\cite{PS}.

The above discussion shows that the Doppler friction factor given by Braginsky and Saulson emerges naturally in a Hamiltonian formalism as long as we avoid the approximation of frequency independent optomechanical coupling strength. Thus Doppler dissipation is a general phenomenon in interferometers. We now want to give a more intuitive explanation of this dissipation within the quantum phonon-photon
scattering picture shown in Fig.2.

\subsection{Sideband photon generation rate}
A more transparent way to investigate our toy model is to calculate the sideband photon generation rate explicitly.
When the external GW force with frequency $\Omega_{\rm gw}$ drives the motion of the test mass (or the
end mirror in our toy model) the optical fields in
the sideband $\omega_0\pm\Omega_{\rm gw}$ appear.
The sideband photon generation rates $R_{\omega_0\pm\Omega_{\rm gw}}(t)$ are given by:
\begin{equation}
R_{\omega_0\pm\Omega_{\rm gw}}(t)=\langle i(t)|\hat{a}^{\dagger}_{\omega_0\pm\Omega_{\rm gw}}(t)\hat{a}_{\omega_0\pm\Omega_{\rm gw}}(t) |i(t)\rangle.
\end{equation}
Here $|i(t)\rangle$ represents the sideband photon states.
For the sidebands with initial vacuum states, we have:
\begin{equation}
|i(t)\rangle=\frac{1}{i\hbar}\int^{t}_0\hat{H}_{int}(t^{\prime})dt^{\prime}|0\rangle.
\end{equation}
Here, the $H_{int}$ is given in (2). Substituting into (7),
after some simple algebra, integrating out all the $\delta-$funtions and take the average over one cycle $2\pi/\Omega_{\rm gw}$ we have:
\begin{equation}
\begin{split}
R_\omega=\frac{E_0^2}{c^2}\omega_0\omega x^2=\frac{P}{\hbar c^2}\omega x^2,
\end{split}
\end{equation}
where $x$ is the amplitude of harmonic motion.
Substituting the two sideband frequencies, we obtain the difference of $\omega_0\pm\Omega_{\rm gw}$ sideband photon generation rates are:
\begin{equation}
R_{\omega_0+\Omega_{\rm gw}}-R_{\omega_0-\Omega_{\rm gw}}=2\frac{P\Omega_{\rm gw}}{\hbar c^2}x^2.
\end{equation}
This is also the mechanical dissipation rate according to particle number conservation.
Clearly, the $\omega_0\pm\Omega_{\rm gw}$ sideband-photon generation rates are only balanced under the approximation $\omega=\omega_0$. Thus Doppler friction effect can be explained as the
result of unbalance between the Stokes and anti-Stokes process rates due to the frequency dependence of the optomechanical coupling constant ($\propto\sqrt{\omega\omega_0}$).

Multipling Eq. (10) by unit phonon
energy $\hbar\Omega_{\rm gw}$, we can express the mechanical power dissipated averaged over one cycle from the test mass as:
\begin{equation}
\begin{split}
\mathcal{P}^m_{diss}=\hbar\Omega_{\rm gw} [R_{\omega_0+\Omega_{\rm gw}}-R_{\omega_0-\Omega_{\rm gw}}]=2\hbar\frac{P\Omega_{\rm gw}^2}{\hbar c^2}x^2=2\frac{\Omega_{\rm gw}^2}{c^2}x^2.
\end{split}
\end{equation}

Expressing the sideband photon generation rate $R_{\omega_0\pm\Omega_{\rm gw}}$ as the generated sideband power over the energy of a single sideband photon: $W_{\omega_0\pm\Omega_{\rm gw}}/(\omega_0\pm\Omega_{\rm gw})$, we can express Eq.(11) as:
\be
\frac{W_{\omega_0+\Omega_{\rm gw}}}{\omega_0+\Omega_{\rm gw}}-\frac{W_{\omega_0-\Omega_{\rm gw}}}{\omega_0-\Omega_{\rm gw}}
=\frac{\mathcal{P}^m_{diss}}{\Omega_{\rm gw}},
\ee
This is the classical form for the Manley-Rowe equation that was originally derived\,\cite{MR} to describe a lossless parametric amplifer using electrical circuit theory. In Appendix B, for completeness, we give the formal derivation of the Manley-Rowe equations using a Hamiltonian formalism.

To compare our power dissipation result with the classical derivation, we recall that the frictional power dissipated is given by $F_vv$. Then using $x(t)=hL\sin(\Omega_{\rm gw}t)$, we can substitute in Eq.(6) and averaged over one cycle to obtain:
\begin{equation}
\langle P_{v}\rangle=-\frac{2P\Omega_{\rm gw}^2}{c^2} h^2L^2.
\end{equation}

This result exactly matches with (12). This can be seen as a result of energy conservation: the energy flow out of the test mass should flow into the optical field. That is: $\langle P_v+\mathcal{P}^m_{diss}\rangle=0$.

The above calculation carries over to conventional interferometers. The detectors commonly considered to have balanced sidebands are\,\cite{PS}, in reality, not precisely balanced. As already emphasised, the imbalance arises because of the frequency-dependence of the optomechanical coupling constant, and is the cause of Doppler friction.

In interferometers, sideband fields carry the gravitational wave information. The sidebands leak into the dark port of the interferometer and are measured by photo-detectors. Each sideband contains the usual displacement dependent term, plus a velocity dependent Doppler friction component. The total power of these sideband optical fields $P_{\text{total}}=P_++P_-$ is given by:
\begin{equation}
\begin{split}
P_{\text{total}}=\frac{2P}{c^2}\omega_0^2x(t)^2+\frac{2P}{c^2}\Omega_{\rm gw}^2x(t)^2,
\end{split}
\end{equation}
where $P$ is the total circulating power.
Both of the sideband terms and Doppler friction term are needed to describe the output
of the interferometer. From this analysis it is clear that the sideband power is not amplified Doppler friction power as previous analysis suggested~\cite{PS}.

\section{Energy absorption in general interferometer configurations}
So far we have discussed Doppler friction in a simple but fundamental
light-mirror interaction model. The energy absorption through Doppler friction is extremely small even when arm cavities like those used in LIGO type detectors are used to enhance the intracavity power.  However, in more general interferometer configurations the energy absorption can be much larger.
Many new interferometer configurations have been proposed, mainly to allow the free-mass standard quantum limit to be beaten through modifying the dynamics of test masses by opto-mechanical interaction.~\cite{DK,BC,BG}.
In the following discussion we will show that these configurations
actually increase the energy absorption from the gravitational wave signal through the creation of unbalanced sidebands. This is achieved by detuning the laser frequency with respect to the resonance of interferometer. We note that although this detuning induced sideband unbalance is different from the unbalance that  occurs in Doppler friction in terms of tunability, it can be understood on equal footing as arising from a variation of density of electromagnetic field modes with respect to frequency.

To formulate the problem, we start from the basic
structure of a general tunable interferometer configuration consisting of an optical cavity with a movable end mirror. For example, the differential mode of a signal recycling laser interferometer operating on the dark port shown in Fig.3(a) can be mapped to a detuned cavity given in Fig.3(b). This mapping relation, shown in (Fig.3) was exactly proved by Buonanno and Chen~\cite{BC} through treating the signal recycling cavity as an effective mirror. Variation of the tuning of the signal recycling mirror accommodates a range of interferometer configurations that includes as a special case the one considered in ~\cite{PS}.
\begin{figure}
\begin{center}
\epsfig{file=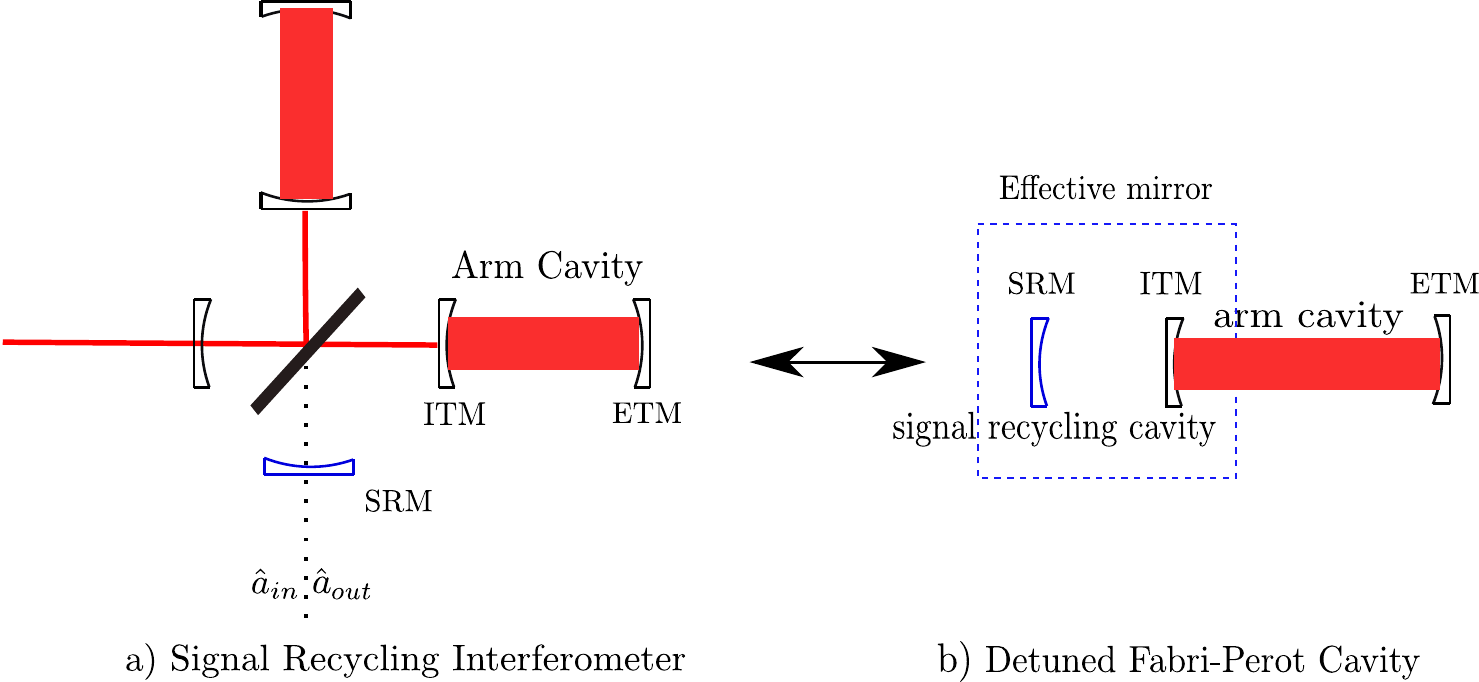,width=0.9\textwidth,height=0.4\textwidth,angle=0}
\caption{A laser interferometer gravitational wave detector, shown schematically in (a) can be treated using a simplified model consisting of a 3-mirror cavity shown in (b).
In (a), ITM indicates input test mass, ETM is the end test mass, while PRM and SRM are the power and signal recycling mirrors.
The solid lines represent the pumping field, while the dotted lines
represent the differential mode which carries the gravitational
signal. The mapping ~\cite{BC} in (b) represents a signal recycling interferometer. The signal-recycling cavity acts as an effective mirror. The position of the SRM determines the detuning. For a
continuous monochromatic gravitational wave signal, the carrier light,
test mass and the two sidebands can be treated as four separate oscillators}
\label{Fig0}
\end{center}
\end{figure}

We will calculate the sideband photon generation rate and its
associated mechanical damping factor which may be positive or negative.
The effective cavity in Fig.3 has a spectral profile such as the one shown in Fig.4. Detuning of the laser frequency from the cavity resonance leads to unbalanced sidebands. The feedback loop
diagram  shown in Fig.5 explains how the detuning creates positive or negative damping. A monochromatic gravitational wave at frequency $\Omega$ acting on the
interferometer causes the test mass to oscillate. The
modulation generates Stokes and anti-Stokes sidebands
inside the system. Both sidebands beat with the main
beam and induce an AC radiation pressure force. However, the Stokes sideband
radiation pressure is in phase with the velocity of mechanical
motion, while the anti-Stokes radiation pressure has  $\pi$ phase shift relative to the velocity of mechanical motion. Thus the Stokes sideband represents positive feedback and can cause
heating effect in which the optical energy will be pumped into the test mass motion and create a tiny increase of gravitational wave strength, while the anti-Stokes causes cooling (or damping). When an
interferometer is unbalanced, its Stokes
sideband becomes higher than its anti-Stokes sideband. This causes the net feedback driving of mechanical modes to be non-zero. By changing the detuning, we change the relative strength of cooling and heating radiation pressure forces.

 To analyze the system quantitatively, we start by writing the Hamiltonian of the system shown in Fig.3(b) in the reference frame of the beam splitter in terms of the optomechanical coupling constant $G_0=\omega_0/L$ and the cavity
 bandwidth $\gamma$. The bandwidth $\gamma$ is given by $cT/4L$, where $T$ is the transmission of the input mirror, $L$ is the cavity length and $c$ is the
 speed of light. In the Hamiltonian, $\hat{a}$ and $\hat{a}_{in}$ are the annihilation operators for the cavity
field and the pumping field, while $\hat{p}$ and $\hat{x}$ are the momentum and displacement operators for the test mass. The frequencies
$\Omega$, $\omega_c$ and $\omega_0$ are the oscillation frequency of test mass motion which is
the gravitational wave frequency, the cavity resonant frequency, and frequency of the pumping light respectively.
Here, let us first put the tiny Doppler friction effect aside, and focus on the optical damping.  In this case it is valid to impose the approximation that optomechanical coupling
strength is a frequency-independent quantity.

\begin{figure}
\begin{center}
\epsfig{file=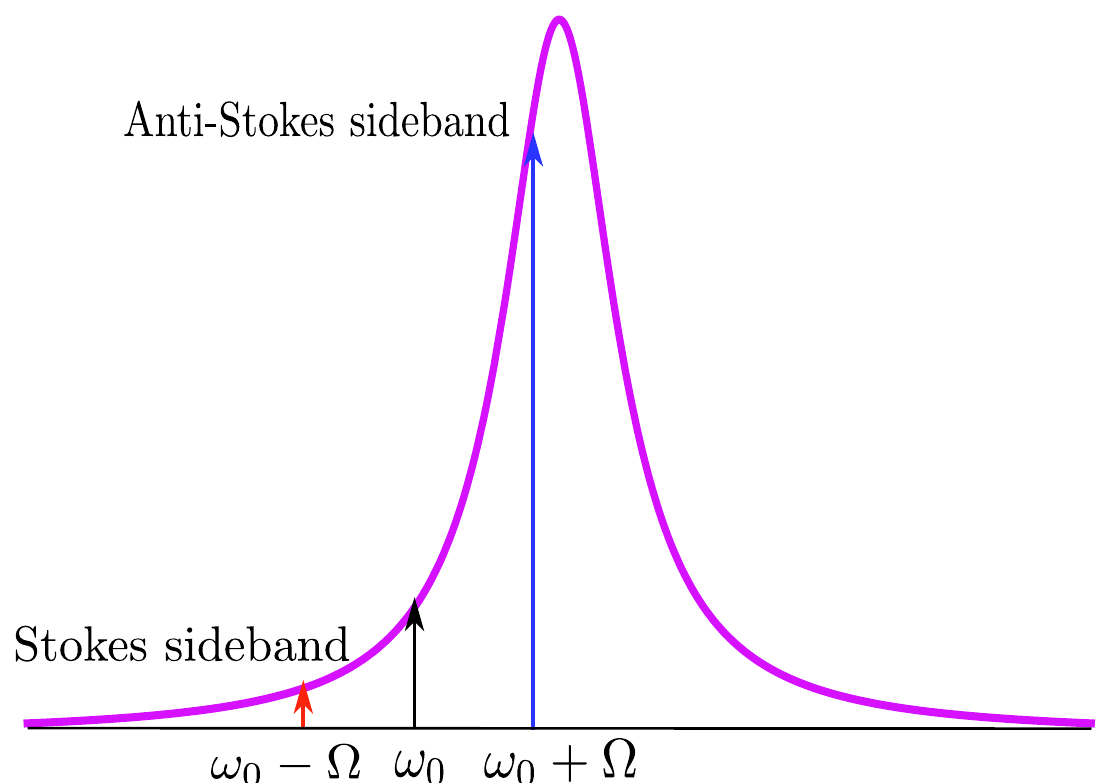,width=0.45\textwidth,height=0.3\textwidth,angle=0}
\caption{Unbalanced cavity spectral profile in the case of
blue detuning. The black, red and blue arrows are the injection
beam, Stokes sideband light and anti-Stokes sideband light
respectively. The pink curve is the cavity spectral profile. The Stokes and anti-Stokes
sidebands are created with different amplitudes due to the frequency dependence of the cavity response.}
\label{Fig0}
\end{center}
\end{figure}

 \begin{figure}
\begin{center}
\epsfig{file=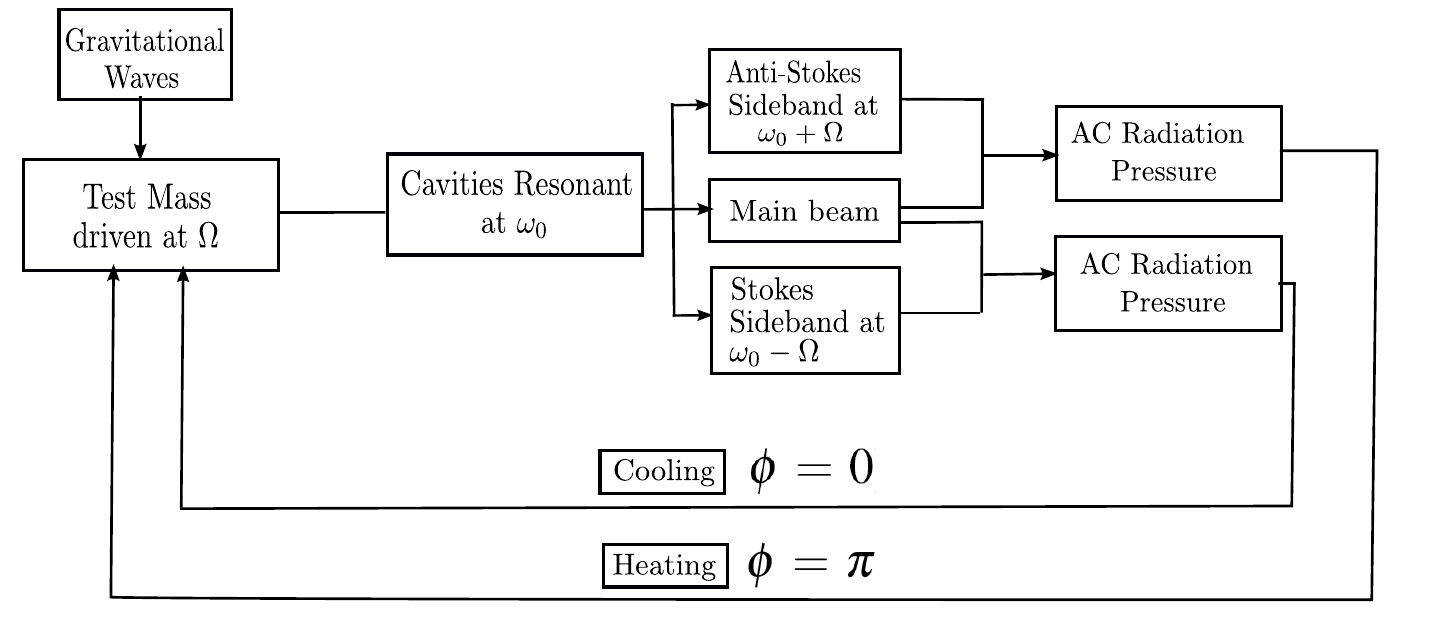,width=0.8\textwidth,height=0.35\textwidth,angle=0}
\caption{Flow chart showing radiation feedback
effects: a gravitational wave modulates the cavity field by driving the motion of the test mass, creating Stokes and anti-Stokes sidebands. The beating of these sidebands with the main laser beam creates a radiation pressure force which acts back on the test mass. The radiation pressure force differs by a phase of $\pi$ for the  anti-Stokes and Stokes sidebands so that one contributes to the cooling and the other to the heating of the test mass motion}
\label{Fig0}
\end{center}
\end{figure}

 The Hamiltonian can be written as~\cite{HSH}:
\begin{equation}
\begin{aligned}
 H=\hbar\omega_c \hat{a}^{\dagger}\hat{a}+p^2/2m+\hbar G_0x\hat{a}^{\dagger}\hat{a}+i\hbar\sqrt{2\gamma}(\hat{a}_{in}
 \hat{a}^{\dagger}e^{-i\omega_0t}-h.c.)-F_{GW}\cdot x.
 \end{aligned}
 \end{equation}
 Here, $-F_{GW}\cdot x$ is the work done by a gravitational wave tidal force on the test mass and $F_{GW}=(1/2)\ddot{h}L$.
 From the above Hamiltonian, we obtain the linearized equations of motion:
\begin{subequations}
\begin{align}
m\ddot{x}(t)=-\hbar
\bar{G}_0[\hat a^{\dagger}(t)+\hat a(t)]+F_{GW}(t),\\
\dot{\hat a}(t)+(\gamma-i\Delta)\hat a(t)=-i\bar{G}_0 x(t)+\sqrt{2\gamma}\hat a_{in}(t).
\end{align}
\end{subequations}
Here, $\bar{G}_0=G_0\bar{a}$ while $\bar{a}$ is the steady amplitude of
the cavity mode and $\Delta$ is the detuning factor defined as $\Delta=\omega_0-\omega_c$.
This detuning can be experimentally realized by tuning the reflectivity and phase
of the signal-recycling mirror.

Taking a Fourier transform of the above equations, we obtain
the following relations:
\begin{subequations}
\begin{align}
 m\Omega^2x(\Omega)=\hbar\bar{G}_0(\hat{a}(\Omega)+\hat{a}^{\dagger}(\Omega))-F_{GW}(\Omega),\\
 \hat{a}(\omega)=\frac{\bar{G}_0x(\omega)}{\omega+\Delta+i\gamma}+\frac{i\sqrt{2\gamma}a_{in}(\omega)}{\omega+\Delta+i\gamma}.
\end{align}
\end{subequations}
The feedback processes shown in Fig.5 are described by Eq.17(a) and (b). The first term on the right hand side of Eq.17(b) describes the effect of the external force driven mechanical motion on the
optical field which in turn is fed back to the mechanical motion
through the radiation pressure force given by the first term on the right hand side of Eq.17(a).

Now we can derive the sideband photon generation rate using perturbation methods. We divide the Hamiltonian into two parts: a) the unperturbed
part consisting of the optical cavity and the mechanical oscillator, and b) the interaction term
$\hbar G_0x\hat{a}^{\dagger}\hat{a}$ representing the perturbed
part. We use the Fermi Golden Rule, which states that the transition rate is proportional to
the square of expectation value of perturbed Hamiltonian. Then we follow a method given by Marquardt et.al~\cite{MC}. We define the (anti-)Stokes process rate
as $R^{\text{(a)S}}$.
Since the mechanical (anti-)damping rate $\Gamma^{\text{aS}}$ measures the relative mechanical energy (gain) loss per unit time ($dE_m/E_mdt$) where $dE_m/dt=\hbar\Omega R^{\text{aS}}$, it follows that:
$$\Gamma^{S}=-\frac{\hbar\Omega R^{S}}{E_m}.$$
The energy change per unit time is just the unit phonon energy times the rate
of the Stokes process. The same applies for anti-Stokes process.
Then we have:
\begin{equation}
 \Gamma^{S}=\frac{\hbar\Omega R^{S}}{E_{m}}=\frac{\hbar\Omega}{\hbar\Omega\bar{n}_m}\frac{(\hbar \bar{G}_0)^2}{\hbar^2}|\langle f|x|i\rangle|^2\langle
 \hat a\hat a^{\dagger}\rangle_{\omega=\Omega}=\bar{G}_0^2\frac{\hbar}{2m\Omega}
 \frac{2\gamma}{(\Omega-\Delta) ^2+\gamma ^2}.
\end{equation}
In deriving this formula, we should substitute (17b) into
$\langle aa^{\dagger}\rangle$ and the free evolution solution
of (17a) into $|\langle f|x|i\rangle|^2$.

For the anti-Stokes process:
\begin{equation}
 \Gamma^{\text{aS}}=\frac{\hbar\Omega R^{\text{aS}}}{E_m}=\frac{\hbar\Omega}{\hbar\Omega\bar{n}_m}\frac{(\hbar \bar{G}_0)^2}{\hbar^2}|\langle f|x|i\rangle|^2\langle \hat a\hat a^{\dagger}\rangle_{\omega=
 -\Omega}=\bar{G}_0^2\frac{\hbar}{2m\Omega}
 \frac{2\gamma}{(\Omega+\Delta) ^2+\gamma ^2}.
\end{equation}

According to the particle conservation law, we have:
\begin{equation}
\Gamma^{m}=\Gamma^{\text{aS}}-\Gamma^{S}.
\end{equation}
This tells us that the mechanical damping rate $\Gamma^{m}$ is given by the difference between Eq.(18) and Eq.(19), which can be simplified to:
\begin{equation}
\Gamma^{m}=-\bar{G}_0^2\frac{\hbar}{m}\frac{2\Delta\gamma}{[(\Omega-\Delta)^2+\gamma ^2][(\Omega+\Delta) ^2+\gamma ^2]}.
\end{equation}
This result is equivalent to the optical damping factor given as the imaginary part of optical rigidity in~\cite{BC,HSH}.  For a typical interferometer cavity used in gravitational wave detection, we plot the form of $\Gamma^{m}$ as a function of cavity detuning $\Delta$ in Fig.6. When $\Delta>0$, the optical damping factor is positive and the radiation pressure force fed back to mechanical motion has the form of $F=m|\Gamma^{m}|\Omega x(\Omega)$. This force is in-phase with the velocity of mechanical motion, and induces the heating effect shown in Fig.(5). However when the $\Delta<0$ , the radiation pressure force has the form of $F=-m|\Gamma^{m}|\Omega x(\Omega)$. It  differs by a $\pi$ phase shift, thereby driving the mechanical motion in anti-phase which induces cooling.
When there is no detuning (i.e. $\Delta=0$), then the transition
rates become equal such that the total optomechanical damping rate is zero:
\begin{equation}
 \Gamma^{S}=\Gamma^{aS}=\bar{G}_0^2\frac{\hbar}{2m\Omega}
\frac{2\gamma}{\Omega^2+\gamma ^2}.
\end{equation}
In this case the anti-Stokes sideband and Stokes-sideband rates are exactly
canceled with each other and $\Gamma^{m}=0$ (under the frequency-independent coupling strength
approximation). This corresponds to the case illustrated in Fig.5, in which the cooling and heating terms cancel each other. Under these circumstances the test masses can be treated as free masses except for the negligible Doppler friction. These results are not only correct for the near-resonance case such as the one shown in Fig.4, but also correct for more general cavity field structures such as the case discussed further below in which a single sideband is resolved and resonant with a high order mode. An analogous single sideband device that manifests the above behavior has recently been experimentally demonstrated by Chen.et.al \cite{Xu}.
\begin{figure}
\begin{center}
  \includegraphics[width=0.8\textwidth]{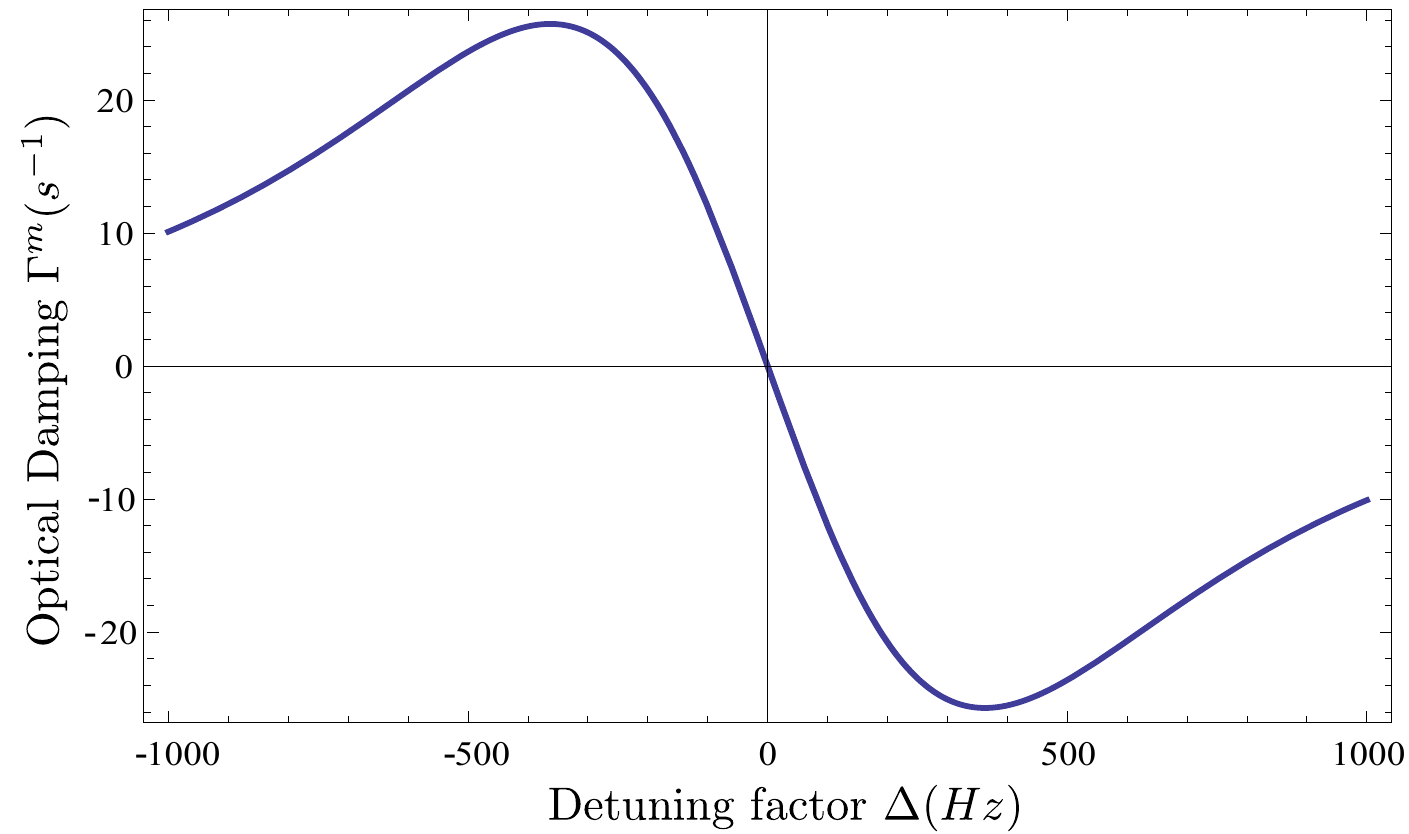}\\
  \caption{The relationship between the optical damping rate $\Gamma^{m}$ and the detuning
  $\Delta$. When $\Delta<0$, the optical damping is positive and corresponds to optomechanical cooling while
  $\Delta>0$, the optical damping is negative and corresponds to optomechanical heating. Here, we take the typical
  interferometer cavity parameters: cavity bandwidth is $100\text{Hz}$, mirror mass is 40kg, intracavity photon number is
  $10^{20}$ and arm length $\sim 4\text{km}$}
\end{center}
\end{figure}

Note that the optical damping factor $\Gamma^{m}$ is always associated
with the optical spring effect, and for a system with optical-damping,
the associated optical spring constant is always negative and hence can lead to
 instability. These relations were discussed by Buonanno et.al~\cite{BC}. However, this instability problem can be solved using the double-optical spring configuration proposed by Rehbein et.al~\cite{RM} or by an electronic feedback loop as proposed by Buonanno et.al~\cite{BC}. Once stabilised, an optical spring interferometer operates like a resonant mass gravitational wave detector. The mechanical stiffness of this detector is contributed by the optical field.

 For the optical spring interferometer, we can calculate the energy in the gravitational wave detector following results already derived long ago by Misner Thorne and Wheeler ~\cite{MTW} for resonant bar detectors. The steady state
vibration energy of the test masses is given by:
\begin{equation}
E_{kin}=\frac{mL^2h^2\Omega_{\rm gw}^6}{16[(\Omega_{\rm gw}^2-\omega_{\text{opt}}^2)^2+\Omega_{\rm gw}^2\Gamma_{\text{eff}}^2]}.
\end{equation}
 Here, the interferometer is treated as a mechanical quadrupole resonator with a resonant frequency $\omega_{\text{opt}}$ due to optical rigidity and $L$, $h$ and $\Omega_{\rm gw}$ are the length of the arm cavity, strain and frequency of the gravitational wave.
The effective test mass damping rate $\Gamma_{\text{eff}}$ is the resonator bandwidth.  For the double optical spring interferometer the damping is contributed by the superposition of two optical springs ( $\Gamma_{\text{eff}}=\Gamma^{m}_1+\Gamma^{m}_2$) \,\cite{RM}. For a feedback stabilised optical spring interferometer the damping is contributed by the sum of optical anti-damping factor and damping rate contribute by the electronic feedback loop  ($\Gamma_{\text{eff}}=\Gamma^{m}+\Gamma^{\text{feedback}}$) \,\cite{BC}.

The steady state kinetic energy (assuming a continuous gravitational wave source) is  dissipated internally at
 a rate $E_{kin}\cdot \Gamma^{\text{eff}}$, which is the energy absorption rate of the detector, or the average absorbed power from the gravitational waves by the detector~\cite{MTW}. The average absorbed power, derived from Eq.22 is shown in Fig.7 for a sinusoidal gravitational wave of amplitude $10^{-23}$ in a typical advanced interferometer. From this, we can see that when the gravitational wave frequency is resonant with the mechanical resonant frequency of the mass-spring system,
 the absorbed power is much higher. It can be  $10^{15}$ times higher than the
 Doppler friction power. Since the energy absorption cross section $\sigma=E_{kin}\cdot \Gamma^{\text{eff}}/\mathcal{F}$ with $\mathcal{F}$ is the gravitational
 wave energy flux, the $\sigma$ is also relatively high ($\sim10^{-22}\text{m}^2$) when the mechanical resonant frequency of the interferometer is resonant with the gravitational wave frequency.

 \begin{figure}
\begin{center}
\epsfig{file=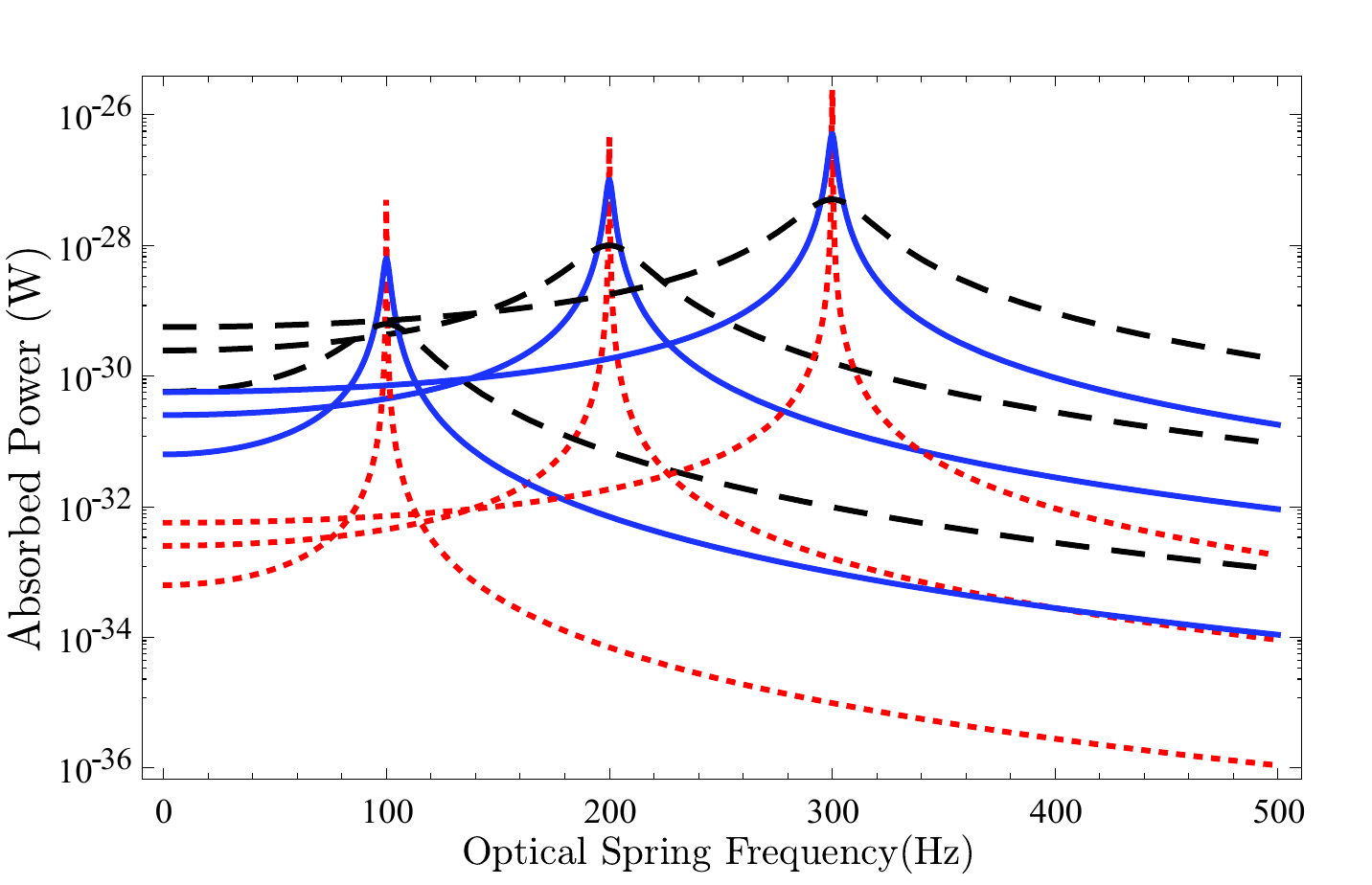,width=0.8\textwidth,height=0.45\textwidth,angle=0}
\caption{The average absorbed power in a double-spring interferometer as a function of optical spring frequency due to monochromatic gravitational waves  of frequency $100$Hz, $200$Hz and $300$Hz and  $h\sim10^{-23}$. We assume a LIGO type interferometer.
The graphs show the absorbed power as a function of optical spring frequency for three different optical damping values. The red-dotted, blue-solid and black-dashed lines represent total optical damping $\Gamma_{\text{eff}}$ of  $0.4$, $40$ and $400$ $\text{s}^{-1}$ respectively.}\label{Fig0}
\end{center}
\end{figure}

We should point out that the Doppler friction power also gets amplified when the detector is resonantly driven by gravitational waves, compared with the power levels given in the last section. The reason is that the Doppler friction power, as shown in Eq.(11), depends on the oscillation amplitude. Estimates show that the Doppler friction power is still $\sim10^{12}-10^{15}$ time smaller, and hence is still negligible.

Since detuning induced sideband imbalance leads to strong energy absorption from gravitational waves, it is interesting to consider other detector designs that can be dominated by a single sideband. One is the tilt interferometer \cite{DGBlair}. Unlike normal interferometers that are designed to detect linear strains, this
configuration is an optical cavity designed to detect tilt motions of the end mirror, which has angular amplitude equal to the strain amplitude h. In this case,
the test mass angular rotation can scatter
the laser field into a $\text{TEM}_{01}$ spatial mode with frequency $\omega_0-\Omega_{\rm gw}$. Due to the asymmetric mode structure of long optical cavities, the upper sideband is suppressed. We present this as an example of a detector that satisfies the requirement of strongly unbalanced sidebands in which the unbalance is achieved without detuning. In practice the tilt interferometer fails to be a significant detector for the LIGO band because its characteristic length is set by mirror size instead of optical cavity length.

\section{Conclusion}
Stimulated by previous work ~\cite{Braginsky}~\cite{PS},
we have been able to obtain a unified understanding of gravitational wave energy absorption by laser interferometers,  combining the intrinsic, but tiny, Doppler friction term with an optical damping term which can be tuned by varying the relative amplitude of the signal sidebands. The Doppler friction itself is explained by
the Stokes and anti-Stokes sideband photon generation rates having a small unbalance caused by the frequency-dependent optomechanical coupling strength.

Our results show that to an excellent approximation, conventional laser interferometer gravitational wave detectors with balanced sidebands can be treated as lossless parametric
transducers in which the energy absorption from
gravitational waves is zero. However in a more general case where detuning or some other technique can cause the sidebands to be unbalanced, variation of the relative strength of the Stokes
and anti-Stokes sidebands leads to strong
optical damping and greatly enhanced absorption of  gravitational wave energy. While our results were derived
for monochromatic gravitational waves, they are true in general because every wave can be
treated as a superposition of monochromatic waves.

The ability to tune the real and imaginary optical spring stiffness through the relative sideband amplitudes is analogous to the variation in in-phase and anti-phase signal feedback used in electronic operational amplifiers, where variation of the feedback is used to change the gain and the input impedance of the amplifier. The design of gravitational wave detectors can be considered from the same viewpoint. The upper sideband acts as an optomechanical servo to null out the spring due to the lower sideband, canceling the optical spring stiffness. This directly changes the mechanical input impedance of the interferometer. Because of the extremely high gravitational wave impedance of free space $\sim c^3/G$, laser interferometers are always poorly impedance matched to gravitational waves. However by increasing the optical stiffness we increase the detector input impedance, thereby reducing the impedance mismatch with free space and increasing the fraction of absorbed energy.

Whether the results presented here implies any advantage for using gravitational wave detectors with more energy absorption still remains an open question since the benefit of such designs can only be determined by considering the signal to noise ratio. This question is beyond the scope of this paper. However, we point out that two designs considered to date, the double optical spring and the detuned resonant sideband extraction interferometer which are equivalent to resonant bar detector design with optical stiffness, achieve enhanced signal to noise ratio that peaks at the frequency where the energy absorption is maximized.

A logical extension of our results is that a detector with an enhanced lower sideband would emit gravitational wave power, acting like a point scattering source with negative cross-section for incoming gravitational waves so that the total gravitational wave power would be slightly enhanced.

\section{Acknowledgements}
We thank our colleagues for fruitful discussion, especially Haixing Miao, Peter Saulson, Harald L\"{u}ck, Farid. Ya. Khalili, Yanbei Chen,
Huan Yang, Maxim Goryachev, Sergey P Vyachatnin
and Stefan.L.Danilishin.
This work was supported by the Australian Research Council and the Department of Education, Science and Training.

\appendix
\section{Derivation of Doppler friction by Lorentz transformation}
Doppler friction can be derived in several ways. For completeness, in this appendix, we give an exact derivation of Doppler friction
using Lorentz transformations of the electromagnetic wave field. This derivation gives exactly the same result as Eq.(12).

We consider the same toy model shown in Fig.2.
For perfectly conducting surface with boundary position $X=x\cos{\Omega t}$, the expression for the reflective electric field $E_{Rfl}$ can be written as:
\be
E_{ref}(X=x\cos{\Omega_{\rm gw} t},t)=-E_0e^{ik_0X+i\omega_0t}e^{-2ik_0x\cos{\Omega_{\rm gw} t}}.
\ee
It could be more rigorous to take boundary conditions of Maxwell equations in the boundary's rest frame. In the case of inertial mirror motion $x=vt$, imposing this rest frame boundary conditions,
we have:
\be
E_{ref}=-\frac{1-\beta}{1+\beta}E_0e^{i(k_0x+\omega_0 t)\frac{1-\beta}{1+\beta}},
\ee
with $\beta =v/c$. This is exactly what would be expected physically: The reflected wave is Doppler shifted in frequency by $\sim(1-2\beta)$, and the reflected waves's Poynting vector is reduced by $\sim(1-4\beta)$. In this case, the reflected light has lost power, which means the light field does work on the mirror at rate $2c|E_0|^2\beta$. Second, the receding mirror in the laboratory frame leaves a growing path of "stored" beam energy in its wake, effectively absorbing power $2|E_0|^2\beta c$. The factor $(1-\beta)/(1+\beta)$ in the above equation accounts for these losses.

In case of periodic motion at frequency $\Omega$ which is of interest to gravitational wave detector, these power flows, to the order of $\beta$, will average to zero. Thereby expansion to order $\beta^2$ is needed. For slow periodic motion, during the first half-cycle motion, the reflected beam energy passing a fixed reference plane is:
\begin{equation}
U_1=P_0\left(\frac{1-\beta}{1+\beta}\right)\left(\frac{x}{c\beta}+\frac{x}{c}\right),
\end{equation}
during the second half cycle motion when the velocity of the mirror changes direction, it is given by:
\begin{equation}
U_2=P_0\left(\frac{1+\beta}{1-\beta}\right)\left(\frac{x}{c\beta}-\frac{x}{c}\right).
\end{equation}
Then the correct power flow per cycle (the period is equal to $2x/c\beta$) is then:
\begin{equation}
P_{cycle}=\frac{U_1+U_2}{2x/(c\beta)}.
\end{equation}
Substitute Eq.(25) and (26) into Eq.(27), keeping terms to the order $\beta^2$ , and taking the average
over one cycle, we have:
\begin{equation}
P_{cycle}=P_0(1+4\langle\beta^2\rangle_{cycle})=P_0(1+2\frac{\Omega_{\rm gw}^2}{c^2}x^2),
\end{equation}
where $x=hL$.
This result exactly matches Eq.(13)(14) in the main text.

\section{Derivation of Manley-Rowe relation}
Here we give a formal proof of
the Mainley-Rowe equations in the multi-mode coupling system such as laser interferometer.
We assume that the system consists of nonlinear coupled oscillators (modes): the carrier light, the test masses which
 are actually pendulums, down-converted sideband light and
 up-converted sideband light. The Stokes and anti-Stokes
 interactions are described by three-mode interaction terms. The Hamiltonian of this system can be written as:
 $$ H=\frac{1}{2}\sum\limits_{i=1}^4
 \frac{p^2}{m_i}+\frac{1}{2}\sum\limits_{i=1}^4 k_i q_i^2+\lambda q_1
 q_2 q_3 +\lambda q_1 q_2 q_4$$
 Here, $\lambda$ describes the strength of the parametric interaction. The $q_i$ is the generalized coordinate for the $i_{th}$ oscillator. 
 Then the equations of motion will be:
\begin{equation}
\begin{array}{l}
\ddot{q_1}+\omega_1^2 q_1=-\lambda q_2 q_3-\lambda q_2 q_4\\
\ddot{q_2}+\omega_2^2 q_2=-\lambda q_2 q_3-\lambda q_2 q_4\\
\ddot{q_3}+\omega_3^2 q_3=-\lambda q_1q_2\\
\ddot{q_4}+\omega_4^2 q_4=-\lambda q_1 q_2\\
\end{array}
\end{equation}
$\omega_i$ here stands for mass-normalized frequency of each
oscillator respectively. For our system,$\omega_1=\omega_c$ which is the frequency of the main laser; the $\omega_2=\Omega$ is the frequency of mechanical oscillation; the $\omega_{3,4}=\omega_c\pm\Omega$ represent the frequency
of two sideband light. By using the
 slowly-variational amplitude approximation, we have the
 following results for the evolution of the oscillator amplitudes
 $A_i$,
 \begin{equation}
\begin{array}{l}
 \dot{A_1}=-\frac{\lambda A_2 A_3}{4 m\omega_c}sin\varphi-\frac{\lambda
A_2 A_4}{4 m \omega_c} sin\theta,\\
\dot{A_2}=\frac{\lambda A_1 A_3}{4 m \Omega}
sin\varphi-\frac{\lambda A_1 A_4}{4 m \Omega} sin\theta,\\
\dot{A_3}=\frac{\lambda A_1 A_2}{4 m (\omega_c-\Omega)} sin\varphi,\\
\dot{A_4}=\frac{\lambda A_1 A_2}{4 m (\omega_c+\Omega)} sin\theta.
\end{array}
\end{equation}
The $\varphi$ terms here are the phase angles of the complex amplitudes. These amplitude evolution equations will lead to:
\begin{equation}
\begin{array}{l}
\frac{1}{2}\frac{d}{dt}(\omega_c
A_1^2+(\omega_c-\Omega)A_3^2+(\omega_c+\Omega)A_4^2)=0,\\
\frac{1}{2}\frac{d}{dt}(\omega_c
A_2^2-(\omega_c-\Omega)A_3^2+(\omega_c+\Omega)A_4^2)=0.
\end{array}
\end{equation}
Note that above equations relate the time variation of energy in different channels. This is the Manley-Rowe equations.
Substituting the oscillator energy $\mathcal{E}=\omega_i^2 A_i^2/2$ into the above equation, we can obtain the following results:
\begin{subequations}
\begin{equation}
\frac{W_p}{\omega_c}+\frac{W_+}{\omega_+}+\frac{W_-}{\omega_-}=0
\end{equation}
\begin{equation}
\frac{W_s}{\Omega}-\frac{W_-}{\omega_-}+\frac{W_+}{\omega_+}=0
\end{equation}
\end{subequations}
The above equations are the Manley-Rowe equations used in the
literature.  Here,$W_-$ and $W_+$ represent the power (change rate of $\mathcal{E}_{\pm}$) in the lower sideband and upper sideband respectively due to the parametric
interaction. This point was also mentioned in Manley-Rowe paper[6], in which they use the terminology "power flow". Following the same definition, $W_s$ here describe the energy flow into (or out from) the mechanical oscillator in the parametric interaction process, $W_p$ is
the energy change of the carrier light. When $\frac{W_-}{\omega_-}$ and $\frac{W_+}{\omega_-}$
have equal value, the $W_s$ is zero.This means that there is no net energy exchange between the light field and the test masses in the system.
\end{document}